\documentclass[showpacs,showkeys,nofootinbib,amsfonts,amssymb,floatfix
,twocolumn,tightenlines,superscriptaddress,prl]{revtex4-2}
\pdfoutput=1

\usepackage{graphicx}  
\usepackage{mathtools}
\usepackage{amsmath, amssymb, bm,bbm}   
\usepackage{bbold}
\usepackage{xcolor}
\usepackage{natbib}
\usepackage{hyperref}

\usepackage{csvsimple} 

\usepackage[normalem]{ulem} 

\usepackage{algpseudocode}
\algrenewcommand\alglinenumber[1]{\bf\small #1:}

\usepackage{siunitx} 

\usepackage{comment}

\usepackage[qm,braket]{qcircuit}

\newcounter{biburlnumpenalty}
\newcounter{biburlucpenalty}
\newcounter{biburllcpenalty}

\newcommand*\xbar[1]{%
   \hbox{%
     \vbox{%
       \hrule height 0.5pt 
       \kern0.3ex
       \hbox{%
         \kern-0.1em
         \ensuremath{#1}%
         \kern-0.1em
       }%
     }%
   }%
} 

\def\){\right)} 
\def\({\left(} 
\def\]{\right]} 
\def\[{\left[}

\begin{document}

\title{Radiative processes on a quantum computer }

\author{%
Paulo F. Bedaque}
\email{bedaque@umd.edu}
\affiliation{Department of Physics, University of Maryland, 
College Park, MD 20742, USA}

\author{%
Ratna Khadka}
\email{rk973@msstate.edu}

\author{%
Gautam Rupak}
\email{grupak@ccs.msstate.edu}

\author{%
Muhammad Yusf}
\email{mf1478@msstate.edu}
\affiliation{Department of Physics \& Astronomy and HPC$^2$ Center for 
Computational Sciences, Mississippi State
University, Mississippi State, MS 39762, USA}

\begin{abstract}
Radiative processes, where a photon/neutrino is emitted as a result of a collision or decay of a particle, play a central role in atomic, nuclear and particle physics. Their rate is determined by certain off-diagonal matrix elements with different initial and final states. We propose a method to compute them using quantum computers. It relies on a single extra qubit that, in a certain sense, represents the photon/neutrino. The generic formula relating this matrix element to the amplitude and frequency of oscillations of the extra qubit follows simply in the near resonance case. We demonstrate the feasibility of  the method by using it in actual quantum computations and simulations of simple systems.
 \end{abstract}

\maketitle

\emph{Introduction.}---
Inelastic reactions are central to several branches of Physics. They are responsible, for instance, for the production of all chemical elements in the early Universe and inside stars. They are, however, particularly difficult to describe theoretically, much more challenging than, for example, the bound states participating in these collisions. In part, this is due to the fact that 
the numerical methods used in realistic calculations of nuclear/atomic bound states rely on stochastic methods and these methods, in one way or the other, use the evolution in imaginary time to project onto the ground state of the system. 
Scattering, however, is an inherently real-time phenomenon. In addition, stochastic methods, when applied to many-fermion systems using imaginary time techniques frequently have a sign-problem that renders them ineffective. A technique using real-time would be able to bypass both of these problems.

The recent advances in quantum computing hardware offers exactly that possibility. 
However, hardware improvements alone are not enough; algorithms to make such calculations feasible are required. In this work we introduce a general method to calculate 
radiative processes, where there is a photon/neutrino in either  final or initial state.

A quantum computer calculation of either $a(b,\gamma)c$ and/or its inverse $c(\gamma,b)a$ process where $a$, $b$, $c$ are nuclei and $\gamma$ a photon,  proceeds through several steps. Firstly, the  fermionic protons and neutrons states in  3-dimensions have to be encoded on
the quantum register efficiently (see, for instance, the  algorithms~\cite{JWT:1928,BravyiKitaev:2002,VerstraeteCirac:2005}.
Secondly, a realistic Hamiltonian has to be chosen and coded in terms of universal quantum gates (simple examples will be shown below and  general methods are extensively studied ~\cite{van_den_Berg_2020,Tomesh:2021pns,Gui:2020}).
Thirdly, the initial state where the participating clusters are formed, has to be created (for instance, by adiabatic evolution). 
These steps are common to the calculation of static properties. Here, we focus on particularities of radiative capture reactions. One might simulate these reactions by starting from a state where the initial nuclei are widely separated, evolve it according to a realistic interaction (including the coupling to the photon/neutrino) and measuring the result after the final nuclei are formed or separated again. This mimics Nature but requires extremely large volumes and evolution times. An alternative, valid for elastic reactions, is to relate  energy levels of the system in a small volume to the phase shifts using the so-called L{\"u}scher formula ~\cite{Luscher:1990ux}, as is done in lattice QCD. Our proposal starts with the observation that the coupling of photon/neutrinos to nuclei/atoms is typically perturbative and the only dynamically non-trivial element required are certain off-diagonal matrix elements of the form $\langle E_f| O_T|E_i\rangle$, where $|E_i\rangle, |E_f\rangle$ are the initial/final states of the nuclei and $O_T$ are certain operators describing the transitions. Contrary to diagonal matrix elements, the off-diagonal transition matrix elements are not given by the expectation value of any measurement and it is not a priori clear how to measure them. 

Here, we propose a method to measure such matrix elements and, consequently, to compute reaction rates for radiative processes. The basic idea is simple and directly inspired by what actually happens in Nature, but simplified so it becomes practically feasible.
In Nature, the photon $\gamma$ in the radiative capture process  $a(b,\gamma)c$ entangles with the large Hilbert space of the environment. The probability for the photon to return and initiate the breakup reaction  $c(\gamma,b)a$ is insignificant. However, in a simulation on a quantum computer, done in a small box, the reaction will proceed forward and backward $a(b,\gamma)c\leftrightarrow c(\gamma,b)a$ and leads to a Rabi oscillation between initial and final states. The aspects of this oscillation are determined by the same transition matrix element that determines the capture/decay rate. This way, the small volume becomes a useful feature, not a hindrance, for the calculation.
Other simplifications of the realistic process can also be used. The electromagnetic field has an infinite number of modes, each one with a particular energy. In non-relativistic systems the photon wavelengths are much larger than the typical sizes of the systems being studied and, in the case of radiative capture or decay, only photons in a very narrow range of energies significantly contribute to the process. Finally, since we are interested in the emission/capture of one photon, there is no need to describe two or more photon states. Thus, the electromagnetic field can be substituted by one single qubit. Its energy can be tuned at will and, in practice, will be chosen close to resonance so the signal we are after is enhanced. 
We then demonstrate our method in a  simple application to E1 electromagnetic transitions involving a particle in a simple harmonic oscillator (SHO). We also consider an even simpler two-level system and use an actual quantum computer to calculate the transition rates between them due to the coupling to a photon.

\emph{Transition matrix element measurement.}---We propose a general strategy to use quantum computers to measure a transition matrix element of the form $\langle E_f| O_T|E_i\rangle$, where $|E_i\rangle, |E_f\rangle$ are eigenstates of the Hamiltonian $H$ (with energy eigenvalues $E_i$, $E_f$, respectively) and $ O_T$ is a generic operator.  For that, we consider the time evolution given by the Hamiltonian
\begin{align}\label{eq:Hresonance}
H_T= \mathbb{1}\otimes H +\frac{\omega}{2} (\mathbb{1}-Z)\otimes \mathbb{1}+c_0\, X\otimes O_T\, ,
\end{align}
acting on the tensor-product space of an additional qubit and the qubits representing the physical system of interest with $c_0$, a potentially dimensionful coupling.  $X$, $Y$, $Z$ are the usual Pauli matrices, and coupling the photon with a $Y$ gate to $O_T$ would also work.

Intuitively, one can think of the extra qubit as representing one ``photon" cavity mode that can be either empty ($|0\rangle$) or occupied ($|1\rangle$) but not multiply occupied as a real photon mode could be. 
The ``photon" frequency $\omega$ can be tuned arbitrarily; it is actually important to have the freedom in choosing $\omega$ near the resonance regime 
$\omega \approx E_f-E_i$, an important difference between our method and Ref.~\cite{Roggero:2018hrn}.
A photon transition $|0\rangle \leftrightarrow |1\rangle$ is always accompanied by a transition between eigenstates of $H$. The rate of these transitions is set by the coupling $c_0$ and the value of the matrix element $|\langle E_f| O_T|E_i\rangle|$.
As mentioned above, contrary to the case of radiative decay in empty space, the coupling of the quantum system to a single mode leads to oscillations between $|E_i\rangle$ and $|E_f\rangle$ instead of a decay from $|E_i\rangle$ to $|E_f\rangle$. In fact,
the frequency and amplitude of these oscillations can be simply related to the matrix element of interest. 
For that, observe that for ``photon" energies $\omega$ near resonance ($\omega \approx E_f-E_i$) the contribution of other eigenstates of the Hamiltonian are suppressed and only two states are relevant: $|1\rangle \otimes |E_i\rangle$ and $|0\rangle \otimes |E_f\rangle$ as they are the only states with total energy near $E_f$. The time evolution of the system is approximately governed by the Hamiltonian (in the basis formed by $|0\rangle \otimes |E_f\rangle$ and $|1\rangle \otimes |E_i\rangle$): 
\begin{align}\label{eq:Heff}
    H^\text{eff}_T =
    \begin{pmatrix}
    E_f & \omega_1/2 \\
    \omega_1/2   & E_i+\omega
    \end{pmatrix},
\end{align} with $\omega_1=2 c_0 \langle E_f|O_T|E_i\rangle $. Assuming  the initial state to be $ |1\rangle\otimes |E_i\rangle $, the probability of finding the system in state $|0\rangle \otimes |E_f\rangle  $ at a later time $t$ is given by:
\begin{align}\label{eq:Pt}
    P(t)= \frac{\omega_1^2}{(\Delta E-\omega)^2+\omega_1^2}\sin^2\frac{\sqrt{(\Delta E-\omega)^2+\omega_1^2} t}{2}\,,
\end{align}with $\Delta E = E_f-E_i$. 
Both the frequency and amplitude of the oscillations depend on the desired matrix element. This suggests the following algorithm:
\begin{algorithmic}[1]
\item Encode the system and  add an extra ``photon" qubit.
\item Prepare the initial  state as $|1\rangle\otimes|E_i\rangle$.  
\item Evolve initial state for time $t$ with $e^{-i H_T t}$.
\item Measure the photon qubit.
\end{algorithmic}

Repeating this process many times and for several values of $t$ determines $P(t)$ and, by fitting it $\langle E_f| O_T|E_i\rangle$ can be determined. Notice that it is not necessary to know $E_f$ a priori as it can be determined by the fit to Eq.~(\ref{eq:Pt}). Also, the transition probability is sizable even for poorly tuned values of $\omega$. The case where two or more final states are close enough to be of importance can be worked out in a similar manner without altering the basic idea. We will not discussed the preparation of the initial state and assume there is an oracle that creates it.

It should be noted that this method can also be used to create an excited state $|E_f\rangle$ starting from $|E_i\rangle$. The particular excited state reached can be chosen by selecting an operator $O_T$ with the proper quantum numbers and adjusting $\omega$ appropriately. Further, in an actual calculation on a quantum computing hardware, a larger coupling $c_0$ can be chosen to make the probability amplitude larger, and make the oscillation frequency shorter so that measurements can be made at small $t$ before the signal deteriorates significantly.
Notice the flexibility introduced by the tunable ``photon energy" $\omega$. By tuning $\omega$ near the resonance, the Hamiltonian $H_T^\mathrm{eff}$ eigenstates $|0\rangle_x$, $|1\rangle_x$ are completely  orthogonal to the initial state $|1\rangle$. Time evolution then efficiently mixes the $|1\rangle=(|0\rangle_x-|1\rangle_x)/\sqrt{2}$ and  $|0\rangle=(|0\rangle_x+|1\rangle_x)/\sqrt{2}$ states, a mechanism  similar to Nuclear Magnetic Resonance.

\emph{Example 1: E1 electromagnetic transition.}---
As an example of the method described in the previous section let us consider a simple model of an E1 transition matrix element in an one dimensional quantum system.

The physical photon field $\bm{A}(\bm{r})$ that induces E1 transition in 3-dimensions is included in non-relativistic quantum mechanics through gauge invariance. The momentum operator is modified through minimal coupling $\bm{P}\rightarrow \bm{P}+q e \bm{A}$ in the kinetic energy term of the Hamiltonian $ \bm{P}^2/(2m)$ of a free particle of mass $m$ and  charge $qe$. The E1 dipole operator is the term with a single $\bm{A}(\bm{r})$ field: $q e \bm{A}(\bm{r})\cdot \bm{P}/m$. One uses the commutator relation $[\bm{R},H_0]=i\bm{P}/m$ 
between initial and final states, $E_i$ and $E_f$ for example,  that are eigenfunctions of $H$ to replace $ q e \bm{A}(\bm{r})\cdot\bm{P}/m \rightarrow i q e (E_f-E_i)\bm{A}(\bm{r})\cdot\bm{R}$. In the large wavelength limit $\bm{A}(\bm{r})$ can be taken to be  $\bm{r}$ independent and  the relevant transition matrix element for the E1 transition is then 
$\langle E_f |\bm{R} |E_i\rangle$. 

Motivated by this physical picture, in 1-dimension,  we consider the transition operator $O_T=X$ with a coupling $c_0=q e \sigma^2$ where we introduced a mass scale $\sigma$ to associate with the constant long wavelength $\bm{A}$ and factor of $E_f-E_i$.  Discretizing space by a lattice with $L$ sites we arrive at the Hamiltonian, in natural units $\hbar=1=c$:
\begin{multline}\label{eq:E1lattice}
     \hat{H}=-\frac{1}{2\hat{m}}\sum_{l}[\hat{\psi}^\dagger_l\hat{\psi}_{l+1}
    +\hat{\psi}^\dagger_{l+1}\hat{\psi}_l
-2\hat{\psi}^\dagger_l\hat{\psi}_l] +\sum_l \hat{V}_l \hat{\psi}^\dagger_l\hat{\psi}_l \\
+\frac{\omega}{2}(\mathbb{1} -Z) + q e \sigma^2 X\sum_l l \hat{\psi}^\dagger_l\hat{\psi}_l\,,
\end{multline}
where  the second quantized  particle field $\psi_l$ destroys one particle at site $l$ and $\hat{V}_l$ is the (discretized version) of the external potential the particle is subjected to.  All the physical parameters have been scaled by appropriate factors of the lattice spacing $a$ to get a dimensionless quantity denoted by a hat. The term proportional to $c_0=q e \sigma^2$ is included in order to calculate the matrix element $\langle E_f| O_T|E_i\rangle$.

For a  SHO potential, we  take $V_l = m\omega_0^2 l^2/2$. The quantum register is comprised of $L+1$ qubits with the photon in the highest qubit and the lower $L$ qubits representing lattice sites $l=-(L-1)/2, -(L-1)/2+1,\dots, (L-1)/2$ for odd $L$. An even potential results in energy eigenfunctions with definite parity. An $l$th-qubit  equal to $|1\rangle$ indicates a particle on site $l$. 

The time evolution is  implemented  by the second order Lee-Suzuki-Trotter approximation formula~\cite{Suzuki:1991}. 
In particular, the particle hopping term $K$, the potential energy term $V$, the photon energy term $K_\omega$ and the E1 transition term $H_{E1}$ are implemented for a small time step $\Delta t$ by
\begin{multline}
    e^{-i(K+V+K_\omega+H_{E1})\Delta t}=
    e^{-i K{\Delta t}/{2}}e^{-i V{\Delta t}/{2}}e^{-i K_\omega\Delta t/2}\\
    \times e^{-i H_{E1}\Delta t}e^{-i K_\omega{\Delta t}/{2}}e^{-i V\Delta t/2}e^{-i K\Delta t/2}+\mathcal O[(\Delta t)^3]\, .
\end{multline}
Longer time unitary evolutions are obtained from the products of small $\Delta t$ evolutions.

The circuit implementing the hopping term  connecting wave function in neighboring lattice site $l$ and $l+1$ are as shown below  with  $\gamma = -\Delta t/m$ for  Trotter time step $\Delta t$. The lower qubits are on the upper wires. 
\begin{small}
\begin{align}
& & & &\begin{array}{c}
    \Qcircuit @C=0.3em @R=1em {
   \lstick{l} & \ctrl{1} &\gate{R_z(-\frac{\pi}{2})}&\targ &\gate{R_y(-\frac{\gamma}{2})}
    &\targ &\gate{R_y(\frac{\gamma}{2})} & \gate{R_z(\frac{\pi}{2})}&\ctrl{1}&\qw\\
  \lstick{l+1} & \targ &\qw &\ctrl{-1} &\qw &\ctrl{-1}& \qw&\qw &\targ&\qw
    }
    \end{array}\nonumber
\end{align}
\end{small}
We use an open boundary condition for the hopping term.
The diagonal term $\hat\psi_l^\dagger\hat\psi_l/\hat{m}$ in the kinetic energy adds only an overall phase and can be discarded.

The potential energy for the harmonic oscillator is implemented by a position dependent single qubit gate $R_z(\theta_l)$ with $\theta_l= -\hat{m}\hat{\omega}_0^2l^2\Delta \hat{t} /2$. Similarly, the photon energy term is implemented with $R_z(-\omega\Delta t)$ on the highest qubit up to an overall phase.

The E1 transition described in Eq.~(\ref{eq:E1lattice}) is implemented as below with $\theta_l=qe\hat{\sigma}^2l\Delta \hat{t}$.
\begin{small}
\begin{align}
& \begin{array}{c}
    \Qcircuit @C=0.3em @R=1em {
   \lstick{l}  &\targ & \ctrl{1} &\gate{R_z(\frac{\pi}{2})}&\targ &\gate{R_y(-\theta_l)}
    &\targ &\gate{R_y(\theta_l)} & \gate{R_z(-\frac{\pi}{2})}&\ctrl{1} &\targ & \qw\\
  \lstick{L} &\ctrl{-1}& \targ &\qw &\ctrl{-1} &\qw &\ctrl{-1}& \qw&\qw &\targ&\ctrl{-1}&\qw
    }
    \end{array}\nonumber
    \hspace{-0.1in}
\end{align}
\end{small}

To calculate the transition probability, we start with the exact ground state and a single photon. It is assumed that we have an oracle to generate the ground state. Then we evolve the system with the full Hamiltonian in Eq.~(\ref{eq:E1lattice}) using second order Lee-Suzuki-Trotter product formula, 
 and take measurements of the photon qubit. The transition probability is the probability of measuring a zero photon. 
 
We simulated this circuit  using the simulator Qulacs~\cite{Suzuki:2021}. We used nuclear physics motivated values $m=\SI{940}{\mega\eV}$,  $\omega_0=\SI{8}{\mega\eV}$, $\sigma=\SI{11}{\mega\eV}$ in the simulations. Lattice spacing $a$ was varied between $\SI{0.5}{\femto\m}$-$\SI{2}{\femto\m}$, keeping the physical volume fixed at $a (L-1)\sim \SI{12}{\femto\m}$. We varied $\Delta t$ to minimize Trotterization error to within 1\%. The exact numerical ground state in a $L$ dimensional Hilbert space of a single particle in a SHO inside a box of size $a(L-1)$ was used for the initial state preparation. 
 
\begin{figure}[htb]
\begin{center}
\includegraphics[width=0.49\textwidth,clip=true]{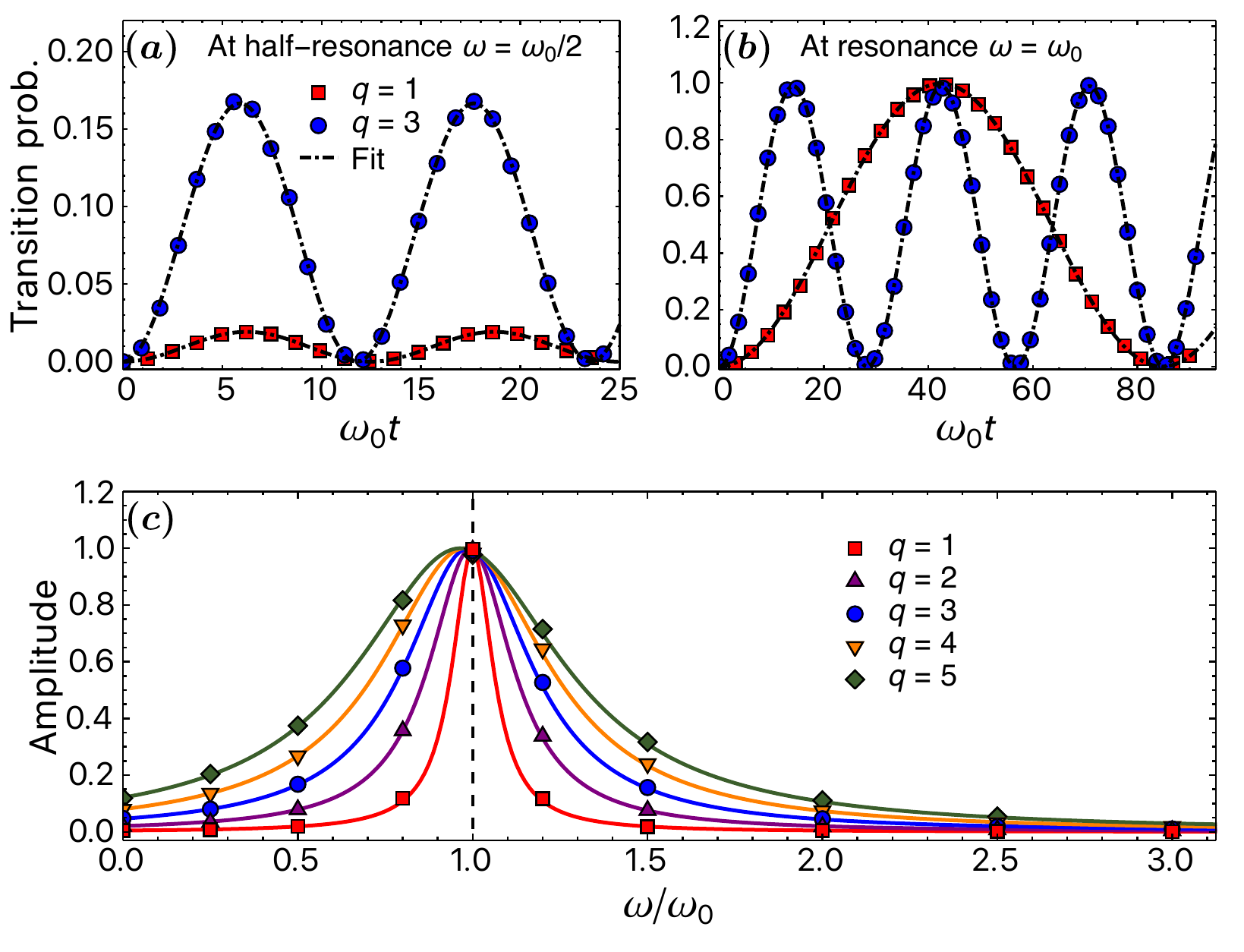}
\vspace{-0.3in}
\end{center}
\caption{\protect  SHO simulation results and fits to probability $P(t)$ in Eq.~(\ref{eq:Pt}) as described in the text.}
\label{fig:SHO}
\end{figure}

 In Fig.~\ref{fig:SHO} we show simulation results  for the SHO with $a=\SI{0.6}{\femto\m}$, $L=21$ and 5000 measurements, and compare with the analytical result for $P(t)$ from Eq.~(\ref{eq:Pt}) with:
 \begin{align}\label{eq:overlapSHO}
    \frac{1}{2}\omega_1^\text{(th)} \equiv q e\sigma^2\int_{-\infty}^\infty dx\, x\phi_1^\ast(x) \phi_0(x)=\frac{qe\sigma^2}{\sqrt{2 m\omega_0}}\,,
\end{align}  where $\phi_n(x)$ are the SHO eigenfunctions for the $n$-th oscillator level. In Fig.~\ref{fig:SHO}, panel $(a)$ are results at $\omega=\omega_0/2$ and panel $(b)$ are results at resonance $\omega=\omega_0$ at $q=1$, 3. Panel $(c)$ compares the probability amplitudes at various $\omega$ and $q$ values to the fit at  $\omega=\omega_0/2$, see table~\ref{table:Rabi}.

In table~\ref{table:Rabi} we fit $\Delta E$ and $\omega_1$ using Eq.~(\ref{eq:Pt}) to the $P(t)$ data at photon frequency $\omega=\omega_0/2$ and resonance $\omega=\omega_0$, through a $\chi^2$ minimization.
The fitted values for  $\Delta E$ and $\omega_1$ agree with the expected values $\omega_0$ and $\omega_1^\text{(th)}$, respectively. The finite volume and discretization errors seem negligible.  

\begin{table}[tbh
]
\centering
\caption{Fit parameter $\Delta\xbar{E}$  and $\xbar{\omega}_1$ for SHO normalized to expected oscillator frequency $\omega_0=\SI{8}{\mega\eV}$ and $\omega_1^\text{(th)}$, respectively. The first set of ($\Delta\xbar{E}$, $\xbar{\omega}_1$) is from fits at $\omega=\omega_0/2$ and the second set with the superscript $R$ is at resonance $\omega=\omega_0$.}
\begin{ruledtabular}
\begin{tabular}{ccccc}
\, $q$ & $\Delta\xbar{E}$& $\xbar{\omega}_1$ & $\Delta\xbar{E}^{(R)}$& $\xbar{\omega}_1^{(R)}$\\ \hline
\begin{filecontents*}{SHO_fits.csv}
q,EA,dEA,OA,dOA,EB,dEB,OB,dOB
1,1.000,0.006,0.936,0.002,0.996,0.008,0.9911,0.0006
2,0.994,0.005,0.967,0.004,0.988,0.011,0.9866,0.0010
3,0.987,0.003,0.972,0.003,0.976,0.008,0.9834,0.0009
4,0.977,0.002,0.966,0.002,0.961,0.010,0.9793,0.0013
5,0.965,0.002,0.958,0.002,0.940,0.010,0.9735,0.0016
\end{filecontents*}
\csvreader[head to column names, late after line=\\]{SHO_fits.csv}{}
{\ \q  
&\num{\EA+-\dEA} & \num{\OA+-\dOA}
&\num{\EB+-\dEB} & \num{\OB+-\dOB}
}
\end{tabular}
\end{ruledtabular}
 \label{table:Rabi}
\end{table}

\emph{Example 2: Two-level system.}---A proof of principle calculation of the algorithm on a physical quantum device is feasible. Consider a two-level system such as the one in Eq.~(\ref{eq:Heff}). It can be thought of as the effective two-level Hamiltonian corresponding to the SHO simulated earlier or as the SHO states in abstract linear vector space instead of coordinate space. However, this identification with SHO is not necessary. The system can be implemented with two qubits.  In Eq.~(\ref{eq:Hresonance}), we represent the two-level system by a single qubit with $H=\omega_0 Z/2$, and $O_T=X$ with SHO coupling $c_0=q e \sigma^2/\sqrt{2 m\omega_0}\equiv\omega_1/2$ (for notational convenience). The two-level Hamiltonian can be any transition operator not necessarily coupled to an external photon. A higher qubit is added for the photon to get: $ H_T = \frac{\omega_0}{2}  \mathbb{1} \otimes Z +\frac{\omega}{2} (\mathbb{1}-Z)\otimes \mathbb{1}+\frac{\omega_1}{2}\, X\otimes X$.

The  $\Delta t$ evolution is implemented by the circuit below in second order Lee-Suzuki-Trotter approximation. 
\begin{small}
\begin{align}
& & &\begin{array}{c}
    \Qcircuit @C=0.3em @R=1em {
   \lstick{0} &\qw &\gate{R_z({\omega_0\Delta t}/{2})}& \ctrl{1} & \gate{R_x(\omega_1\Delta t)} & \qw
   &\ctrl{1} &\qw &\gate{R_z({\omega_0\Delta t}/{2})} &\qw &\qw \\
  \lstick{1} &\qw  &\gate{R_z(-{\omega\Delta t}/{2})}  &\targ &\qw &\qw&\targ &\qw
  &\gate{R_z(-{\omega\Delta t}/{2})} &\qw &\qw
    }
    \end{array}\nonumber
\end{align}
\end{small}

\begin{figure}[tbh]
\begin{center}
\includegraphics[width=0.49\textwidth,clip=true]{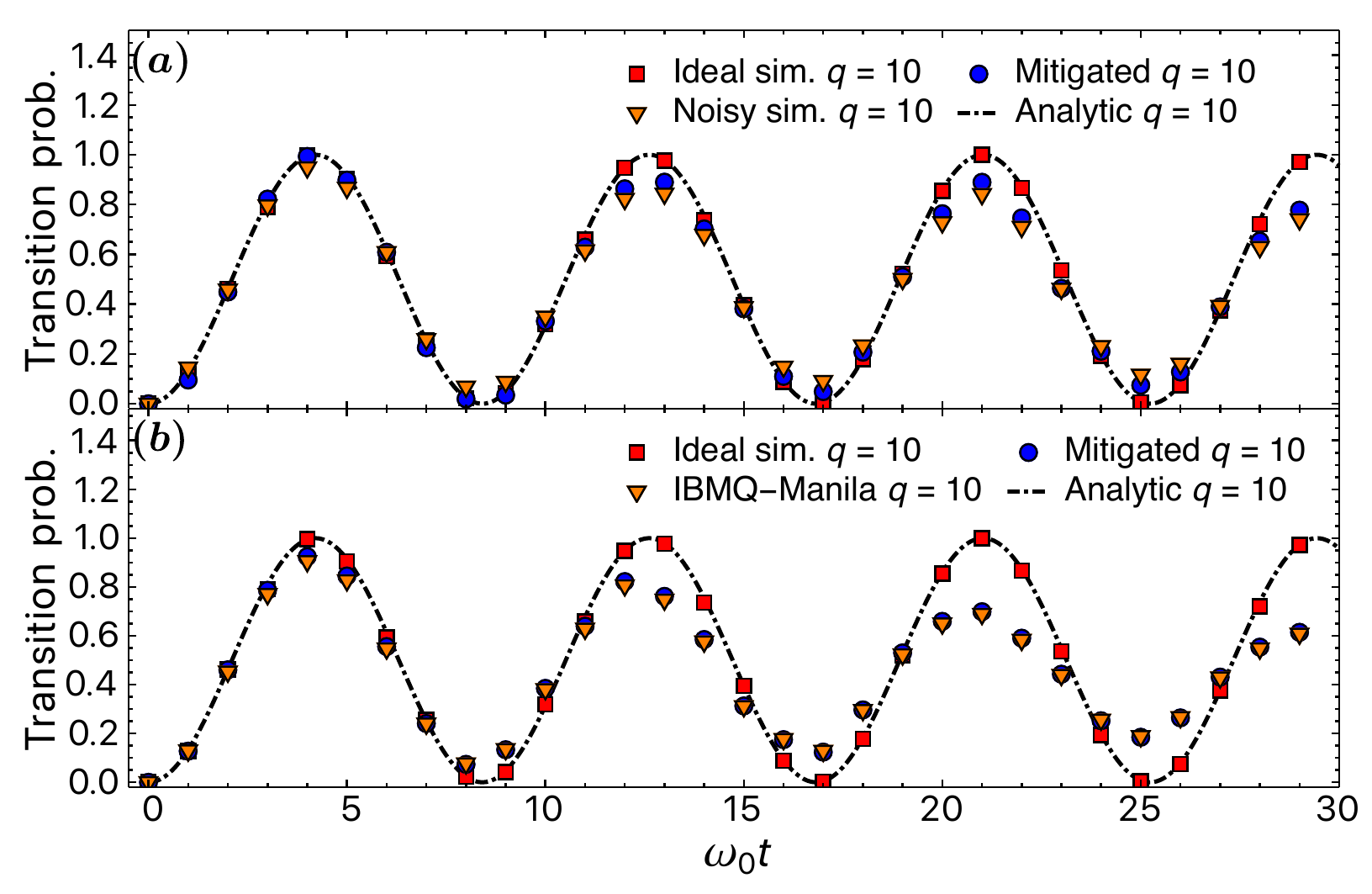}
\vspace{-0.2in}
\end{center}
\caption{\protect Two-level results for ideal simulation, noisy simulation and IBMQ-Manila as described in the text. }
\label{fig:2level}
\end{figure}

A noisy simulation using the qiskit simulator backend FakeManila~\cite{Qiskit} 
is presented in Fig.~\ref{fig:2level}, $(a)$. A large value $q=10$ was used in order to make the oscillations fast and therefore measurable before decoherence sets in. The rest of the parameter values for $\omega_0$, $\sigma$, $m$, etc., were the same as the SHO. A time step $\Delta t = 1/\omega_0$ was small enough for this measurement.  Measurements on the physical qubits 3 and 4 of  ibmq\_manila v1.0.35 (IBMQ-Manila)~\cite{IBMQ-Manila} are shown in Fig.~\ref{fig:2level}, $(b)$. The ideal simulations, noisy simulations and hardware results in Fig.~\ref{fig:2level} were obtained from 1024 repeated measurements. 
We used a simple measurement error mitigation  for the noisy simulations and hardware measurements where the corrected transition probability $P_{c}(t)$ is given by $P_c(t) = M^{-1} P_\text{measured}$, where $M$ is calculated from  measurements on the initial states $|00\rangle, |01\rangle, |10\rangle$ and $|11\rangle$, see Ref.~\cite{ErrorMitigation:2021} and references therein. Error mitigation improves the signal noticeably for the noisy simulation but not for the hardware measurements.  It would seem that even as the signal deteriorates, one can infer at resonance $\omega=\omega_0$, from the amplitude maxima $\omega_1^\text{(meas.)}=(2n+1)\pi/t_\text{max}$ or minima $\omega_1^\text{(meas.)}=2(n+1)\pi/t_\text{min}$ for $n=0,1,\dots$, in Fig.~\ref{fig:2level}. Averaging over the maxima and minima locations, we find $\omega_1^\text{(meas.)}\approx \num{1.03+-0.03}\,\omega_1^\text{(th)}$. 

\emph{Conclusions.}--We developed a general method for calculating matrix elements and cross sections for electroweak transitions. This method can be generalized to charge exchange reactions $a(b,c)d$ by introducing a resonance by coupling to a ``photon" counting register. The algorithm works even if the desired transition operator $O_T$ in Eq.~(\ref{eq:Hresonance}) didn't originally couple with an external photon or neutrino as demonstrated with the two-level result. Our method can be used to calculate excited state wave function, including the spatial dependence,  with 100\% probability, in principle,  near the resonance which is more efficient than a similar method proposed in Refs.~\cite{Roggero:2018hrn,Roggero:2020sgd}. LCU methods~\cite{PhysRevLett.114.090502,ChildsWiebe:2012} to perform the Hamiltonian time evolution can be used for efficiency/complexity reduction. In this work, the Hamiltonian only had discrete levels unlike the continuum states in  real processes such as $d(\gamma,n)p$. However, for calculations in a box, the continuum states are also discrete. A more general method involving the density of states $g(E)$ at continuum energy $E$ near a resonance have to be explored in a future work.

\emph{Acknowledgments.}---This work was supported in part by
U.S. DOE grants DE-SC0021195 (RK, GR, MY), DE-SC0021143,   DE‐FG02‐93ER40762 and DE-SC0021143 (PB) and NSF grants PHY-1913620 (GR, MY), PHY-2209184 (GR).
The  figures for this article have been created using SciDraw~\cite{SciDraw}.

\setcounter{biburlnumpenalty}{9000}
\setcounter{biburllcpenalty}{7000}
\setcounter{biburlucpenalty}{8000}

%


\end{document}